\pdfoutput=1
\documentclass[aps,prl,reprint,groupedaddress,floatfix]{revtex4-1}

\usepackage{graphicx}
\usepackage{dcolumn}
\usepackage{bm}
\usepackage{physics}
\usepackage{amssymb}
\usepackage{amsmath}

\begin{document}


\title{Rapid technique for preparing conductive graphene-borosilicone putty}


\author{Louis A. Bloomfield}
\email{lab3e@virginia.edu}
\affiliation{Department of Physics, University of Virginia, Charlottesville, VA 22904}


\date{\today}

\begin{abstract}
When borosilicone putty is filled with graphene sheets, the result can be a ``g-putty''---an electrically conducting putty that is exquisitely sensitive to deformation. In this work, we present a rapid method for producing g-putty from commercially available ingredients. The process can be performed at room temperature and has methanol as its only waste product.

\end{abstract}


\maketitle

Since their discovery in 1940,\cite{rochow1945} borosilicones have intrigued scientists and entertained children. For the most part, however, borosilicones have been a solution in search of a problem. Similarly, graphene has been fascinating to scientists for decades,\cite{novoselov2004} but it's practical applications outside the laboratory remain limited.

The 2016 discovery that graphene-filling borosilicones are electrically conductive,\cite{boland2016} with conductances that are exquisitely sensitive to deformation due to percolation effects,\cite{marsden2018electrical} suggests many practical sensor applications. However, preparing these ``g-putties'' has required complicated processes that involve solvents or heat or both. In this work, we describe a simple technique for preparing g-putties, one that starts with commodity chemicals and produces an electrically conductive g-putty less than 15 minutes at room temperature.

The ingredients used in our process are:
\begin{enumerate}
	\item Silanol-terminated silicone fluid (liquid)
	\item Trimethyl borate (liquid)
	\item Iso-stearic acid (liquid)
	\item Graphene platelets (fine powder)
\end{enumerate}
The silanol-terminated polydimethylsiloxane (OH-PDMS-OH or STPDMS) fluid used was Dystar Masil SFR 70 (nominal viscosity 70 cSt), but any low-viscosity STPDMS will work. The trimethyl borate used was Alfa-Aesar B20215 trimethyl borate, 99\%, but any similar product will work. The iso-stearic acid used was Nissan Chemical Iso-Stearic Acid-N, but any commercial iso-stearic acid or other non-volatile liquid carboxylic acid (e.g., oleic acid) can be substituted. The graphene platelets used were Alfa-Aesar 47132 graphene nanoplatelets aggregates, sub-micronparticles, S.A. 500 m${}^2$/g. Dis-aggregated or otherwise more highly refined graphene sheets can surely be substituted for the platelets used here. 

The STPDMS fluid contains silicone polymer chains that are covalently crosslinked by the trimethyl borate. That conceptual step is similar to the one used to make ordinary silicone rubbers. But while most ordinary covalent crosslinks between silicone polymer chains form only in the presence of catalysts, elevated temperatures, or both, the covalent crosslinks formed by trimethyl borate requires neither. Instead the boron crosslinks form rapidly at room temperature and the crosslinking process can be driven to completion in seconds under proper circumstances.\cite{bloomfield2018borosilicones} 

\begin{figure}
	\includegraphics{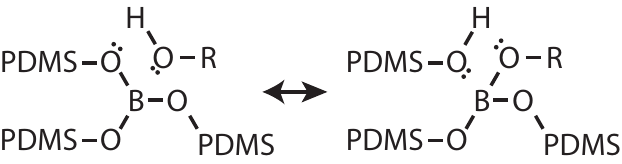}%
	\caption{A proposed mechanism for the exchange reaction between a trimethyl borate molecule ((CH${}_3$O)${}_3$B or TMB) and a hydroxyl-containing molecule (R-OH). Like many simple boron compounds, TMB is planar and its boron atom has an empty p-orbital. That empty p-orbital is easily attacked by a lone electron pair on the R-OH molecule's oxygen atom, resulting in the exchange of ligands. The R-OH molecule becomes an RO- ligand on the boron atom and a CH${}_3$O- ligand leaves the boron atom as a methanol molecule (CH${}_3$OH).\label{fig:exchangeReaction}}
\end{figure}

To explain that rapid crosslinking, as well as the unusual behaviors of borosilicones, we note that trimethyl borate (TMB) undergoes rapid redistribution reactions with other boron esters.\cite{heyes1968} More generally, TMB undergoes rapid ester exchange with molecules bearing -OH groups (hydroxyls or carboxylates). A proposed mechanism for that exchange is shown in Fig. \ref{fig:exchangeReaction}. Because the ester exchange reaction occurs rapidly in both directions, thermodynamic equilibrium is quickly achieved.

To drive the TMB crosslinking reaction to completion, however, the equilibrium must be disturbed in accordance to Le Chatelier's principle. Since a methanol molecule is formed whenever a methoxy group on a TMB molecule is replaced by another ester, removing the volatile methanol molecules from the equilibrium mixture can drive that reaction in one direction so that it reaches completion. As long as enough STPDMS fluid was initially present to replace all of the TMB's methoxy groups, there will be almost zero TMB remaining in the mixture. We note that when Rochow first produced a borosilicone in 1940,\cite{rochow1945} it was as an accident involving TMB. Since then, boric acid has been the more common crosslinker choice, a choice requiring elevated temperatures.

A borosilicone (boron-crosslinked silicone) can be produced by combining any STPDMS fluid with TMB and removing all of the methanol produced by the ester exchange reactions. If that mixture was initially stoichiometrically balanced, with as many moles of silanol (-OH) groups in the STPDMS as moles of methoxy (-OCH${}_3$) groups in the TMB, then removing the methanol will produce a fully crosslinked borosilicone. The methanol can be allowed to evaporate spontaneously from the surface of the mixture, but a faster approach is to apply vacuum to the mixture and cause the methanol to boil out of the mixture. The resulting extra surface area of the methanol vapor bubbles greatly increases the speed at which methanol leaves the mixture.

In the absence of any remaining -OH bearing molecules, a fully crosslinked borosilicone is a network solid---a gigantic macromolecule in which a single covalently-bonded network extends throughout the entire material.\cite{bloomfield2018borosilicones} If -OH bearing molecules are present in the borosilicone, however, the -OH groups will continue to undergo ester exchange reactions on the boron atoms and render the boron crosslinks \emph{temporary} rather than \emph{permanent}. The temporary nature of its crosslinks allows the boron-crosslinked material to relax stress and to flow as a fluid. Consequently, a fully crosslinked borosilicone in which there are -OH bearing molecules is a network liquid---a gigantic macro-molecule in which a single covalently-bonded network extends throughout the material, but with covalent bonds that detach and re-attach frequently so that the network can evolve in topology and geometry.

It is impossible to remove all -OH bearing molecules from a borosilicone, particularly since water and atmospheric moisture participate in the ester exchange reactions. Thus real borosilicones are never quite network solids; they are always network liquids, though perhaps with extremely large viscosities. It is often desirable, however, to lower the viscosity of a borosilicone by adding -OH bearing molecules. For that purpose, carboxylic acids are ideal. The ester exchange reactions involving carboxylate groups (-COOH) are particularly rapid, so a tiny concentration of a carboxylic acid can greatly increase the rate at which boron crosslinks exchange esters. As little as 0.01 wt\% iso-stearic acid (ISA) significantly shortens the average lifetime of a borosilicone's boron crosslinks and thereby lowers its viscosity. Adding a small concentration of ISA is thus necessary to produce a pliable putty rather than a nearly solid material.

The final ingredient in producing a g-putty is graphene itself. The best graphene consists entirely of isolated sheets, each a monolayer thick and as wide as possible along its plane. In this work, graphene platelets are used. Those platelets are much thicker than monolayers and are consequently not ideal for g-putty. They are, however, relatively inexpensive and available, and are used in the present work.

Using the basic concepts discussed above, six putties were prepared, differing only in their graphene concentrations (See Table \ref{tab:putties}). In each case, a measured quantity of graphene powder was added to 10.0g of STPDMS fluid in a 100ml polypropylene beaker and the two ingredients were stirred by hand with a glass rod until well blended, about 30 seconds. To this mixture were added approximately 0.020g ISA (two pipette drops) and 0.250g TMB. The assembled mixture was then blended and kneaded together by immersing a rotating rod (1.27cm diameter stainless steel, rotating vertically at 180 rpm) into the mixture and pressing that rod against the wall of the beaker.

After about 30 seconds of kneading, the beaker of sticky mixture was placed in a vacuum desiccator and subjected to vacuum for about 1 minute to remove the methanol. The spinning rod was used a second time, along with a second minute of vacuum drying. The putty, now virtually free of TMB or methanol, was removed from the beaker and kneaded until homogeneous, using variously a 2-roll hand mill (equivalent to a pasta roller), an arbor press, or gloved hands. The putty was then tested for electrical conductivity. The total time elapsed, from empty beaker to testing for conductivity was less than 15 minutes.

The six putties produced in this fashion are listed in Table \ref{tab:putties}. All are jet black, but only the last four (C, D, E, and F) were found to be electrically conducting g-putties. The other two putties (A and B) had no observable electrical conductivity, so the threshold concentration for electrical conductivity is apparently between 9.1wt\% graphene and 13.0wt\% graphene. 
\begin{table}
	\caption{Graphene-containing borosilicone putties produced by combining STPDMS silicone fluid, graphene platelet powder, iso-stearic acid, and trimethyl borate.\label{tab:putties}}
	\begin{ruledtabular}
		\begin{tabular}{lrrc}
			\textrm{Putty}&
			\textrm{Graphene (g)}&
			\textrm{Graphene (wt\%)}&
			Conductivity (S/m)\\
			\colrule
			A & 0.500 & 4.8 & 0\\
			B & 1.000 & 9.1 & 0\\
			C & 1.500 & 13.0 & $3\times 10^{-10}$\\
			D & 2.000 & 16.7 & $3\times 10^{-8}$\\
			E & 2.500 & 20.0 & $5\times 10^{-6}$\\
			F & 4.000 & 28.6 & $3\times 10^{-4}$
		\end{tabular}
	\end{ruledtabular}
\end{table}

In summary, we demonstrate that g-putty can be prepared from commercially available materials in less than 15 minutes at room temperature with simple equipment. This rapid technique uses trimethyl borate as the crosslinking agent and is facilitated by vacuum removal of the methanol released by the crosslinking reaction. Using commercial graphene platelets, the percolation threshold above which the putty becomes electrically conductive as likely between 9.1wt\% and 13.0wt\% graphene.

\bibliography{graphene-borosilicone}

\end{document}